\documentclass[twocolumn,twoside,slac_two]{revtex4}
\usepackage{graphicx}
\usepackage{fancyhdr}
\def\Fermi{\emph{Fermi}}

\begin{document}

\title{A new model for the \Fermi-LAT Extragalactic Gamma-Ray Background}


\author{M.~Cavadini}
\affiliation{Dipartimento di Scienza e Alta Tecnologia, Universit\'a
  dell'Insubria, via Valleggio 11, I-22100 Como, Italy}
\author{R.~Salvaterra}
\affiliation{INAF/IASF-MI, via Bassini 15, 20133 Milano, Italy}
\author{F.~Haardt}

\affiliation{Dipartimento di Scienza e Alta Tecnologia, Universit\'a
  dell'Insubria, via Valleggio 11, I-22100 Como, Italy}
\affiliation{INFN, Sezione di Milano-Bicocca, I-20126 Milano, Italy}

\begin{abstract}
We compute the contribution of blazars (Flat Spectrum Radio Quasars
and BL-Lacs) to the \Fermi-LAT Extragalactic $\gamma$-ray Background,
assuming a $\gamma$-ray luminosity function proportional to the
luminosity function of radio galaxies \cite{Willott} and the
Spectral Energy Distribution of blazars given by the so-called
\emph{blazar sequence} \cite{Fossati}. We find that blazars account
for the $\sim$45 \% of the total \Fermi-LAT EGB. However, at low
($<$10 GeV) and high ($>$50 GeV) energies the blazar contribution
falls short to account for the observed background light. At E$<$10
GeV, the Extragalactic $\gamma$-ray Background can be fully explained by considering the additional
contribution of star-forming galaxies, whereas at very high energies,
where the absorption due to Extragalactic Background Light dominates, the observed background calls for an additional unknown component.

\end{abstract}

\maketitle

\thispagestyle{fancy}


\section{Introduction}
The Extragalactic $\gamma$-ray background (hereafter EGB) represents a
fascinating challenge since its first detection by \emph{SAS-2}
satellite above 30 MeV \cite{Fic78}. The
$\gamma$-ray telescope \emph{EGRET} during nine years of operations,
measure an isotropic $\gamma$-ray emission in the 30 MeV-30 GeV band
\cite{Strong}.
The satellite \emph{Fermi}-LAT after one years of observations improves the
\emph{EGRET} measurement confirming the existance of an EGB up to 100
GeV \cite{Abdofondo}.\\
As already found by \emph{EGRET} and confirmed by \emph{Fermi}-LAT
\cite{Fermi_coll}, blazars are the most common objects in the
extragalactic $\gamma$-ray sky. Blazars, radio loud AGNs with the
jet pointing towards the Earth, are classified as Flat Spectrum Radio
Quasars (FSRQs) and BL-Lacs. After three years of observations,
\Fermi-LAT detected approximately 1000 blazars with about the same
number of FSRQs and BL-Lacs \cite{Fermi_coll}. It results to be natural computing
their contribution to the EGB (e.g. \cite{Dermer, Inoue, Stecker,
  vent}). The main ingredients in the EGB calculation are the
 $\gamma$-ray luminosity function (LF) of blazars and their spectral energy distribution (SED).
In our model adopt a LF of FRI and FRII radio galaxies \cite{Willott}
and the blazar SEDs using the the so-called \emph{blazar sequence}
\cite{Fossati, Donato}. We compute the blazar
contribution to the \Fermi-LAT EGB fitting the differential \Fermi-LAT
logN-logS, having as a sole fitting parameter the fraction of radio
galaxies beamed towards the Earth, hence appearing as blazars.
Beacause the blazar component alone is not able to fully account for
the EGB at low energies ($<$10 GeV), we consider the $\gamma$-ray
emission of star-forming galaxies. This allows us to fully explain the EGB but the last point at
approximately 80 GeV where the EBL absorption strongly dominates.
Notice that here we refer to EGB as the sum of the resolved source
component and the unresolved and/or truly diffuse emission.

\section{Blazar contribution}

The blazar contribution (in photons s$^{-1}$ cm$^{-2}$ sr$^{-1}$ MeV$^{-1}$) to the EGB at the observed energy $E_0$ is 
%%%%%%%%%%%%%%%%%%%%
\begin{eqnarray}\label{eq.blazars}
I_{\rm{blaz}}(E_{0}) &=& \frac{1}{4\pi} 
\int_0^{\infty} dz\, \frac{dV}{dz}
\int_{\log L_{\gamma}^{\mathrm{min}}}^{\log L_{\gamma}^{\mathrm{max}}} d\log L_\gamma\,
\frac{d\Phi_\gamma (L_\gamma,z)}{d\log L_\gamma} \nonumber\\
&\times&\frac{dn(L_\gamma,z)}{dE} e^{-\tau_{\gamma \gamma}(E_0,z)},
\end{eqnarray}
%%%%%%%%%%%%%%%%%%%%%
where $d\Phi_\gamma(L_\gamma,z)/d\log L_\gamma$ is the $\gamma$-ray LF and 
$L_{\gamma}$ is  $\nu L_\nu$ (in erg/s) at 100 MeV, 
$dn(L_\gamma,z)/dE$ is the unabsorbed photon flux per unit energy $E=E_0(1+z)$
measured on Earth of a blazar with luminosity $L_{\gamma}$ at redshift $z$, and $\tau_{\gamma \gamma}(E_{0},
z)$ is the optical depth for $\gamma-\gamma$ absorption. We adopt the Extragalactic Background Light (EBL) model by \cite{Finke}. 
In the above equation $dV/dz$ is the comoving cosmological volume. 
We set $\log L_{\gamma}^{\mathrm{min}}=43.5$ and $\log L_{\gamma}^{\mathrm{max}}=50$.

The number of sources $N(>F_{\rm ph})$ per steradian with photon flux
greater than $F_{\rm ph}$ is
%%%%%%%%%%%%%%%%%%%%
\begin{equation}
N(>F_{\rm ph})= \frac{1}{4\pi}
  \int_{0}^{\infty}dz\, \frac{dV}{dz} 
 \int_{\log L_{\gamma}^{\mathrm{min}}}^{\log L_{\gamma}^{\mathrm{max}}} d\log L_\gamma\,
 \frac{d\Phi_\gamma(L_\gamma,z)}{d\log L_\gamma}.
\label{eq:counts}
\end{equation} 
%%%%%%%%%%%%%%%%%%%%
The $\gamma$-ray LF of blazars is presently uncertain (see e.g. \cite{AbdoLF}), 
so that one has to rely on the LFs computed in other bands, e.g.,
X-rays \cite{Narumoto, Inoue}, or radio \cite{Stecker}. We adopt here the 
radio LF at 151 MHz of FRI and FRII \cite{Willott}, assumed to be the parent populations of blazars: 
%%%%%%%%%%%%%%%%%%%%%
\begin{equation}\label{blaz}
\frac{\Phi_{\gamma}(L_{\gamma},z)} {d\log L_{\gamma}}=\kappa~\frac{\Phi_{R}(L_{R},z)}{
d\log L_{R}},
\end{equation}
%%%%%%%%%%%%%%%%%%%%%
where $L_{\rm R}$ is $\nu L_\nu$ at 151 MHz, and 
the constant $\kappa$ is the fraction of blazars over all radio galaxies. 
In order to convert radio into 
$\gamma$-ray luminosity, we rely on the blazar spectral energy distribution (SED). 
 As pointed out in \cite{Fossati}, a relation between
the radio and/or bolometric luminosity and the energy of the two peaks exists 
(the so-called {\it blazar sequence}, see also \cite{Donato}). 
We use the SEDs computed by \cite{Inoue, Donato}.

\begin{figure}
\centering
\includegraphics[scale=0.43]{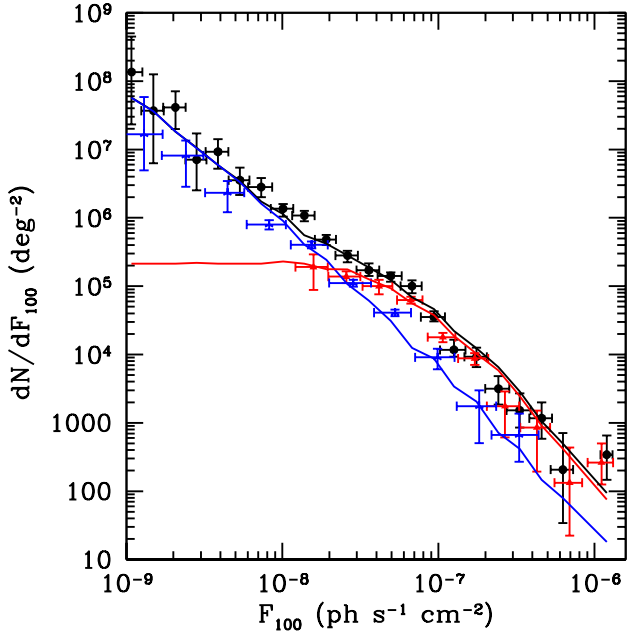}
\caption{The black line represents the fit to the blazar logN- logS  measured 
 by Fermi \cite{Abdocount} as a sum of BL-Lac component (blue line)
 and FSRQ component (red line). Black points are the number counts of
 all the blazars, red points FSRQs and blue points BL-Lacs} \label{fig:conteggi}
\end{figure}

\begin{figure}[t]
\centering
\includegraphics[scale=0.43]{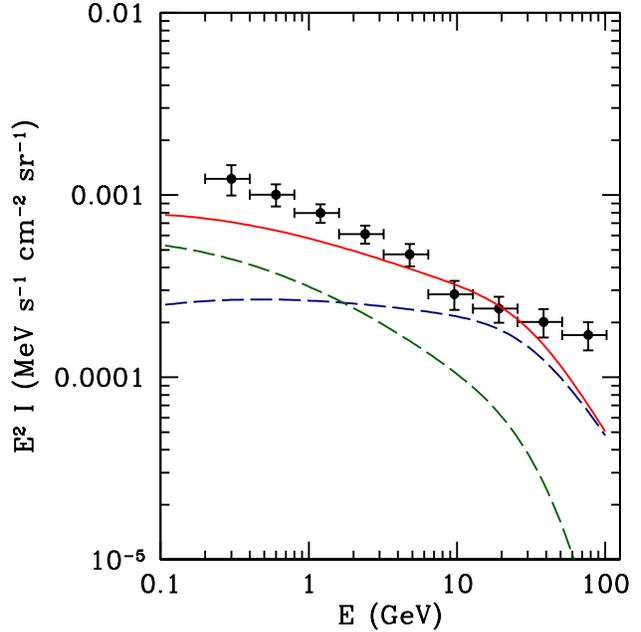}
\caption{The black points represent the total EGB (resolved and
  unresolved sources and all the rest), blue line is  the FSRQ
  component, the green line is the 
BL-Lac component. The red line is the sum of the two components.} \label{fig:paola}
\end{figure}

\subsection{Results}
We use the model discussed in the previous section to compute the
total contribution of blazars (FSRQs and BL-Lacs) to the {\it
  Fermi}-LAT EGB. To this end we fit blazar logN-logS distribution by
letting $k$ to vary. We find $\kappa=(3.93 \pm 0.01) \times
10^{-4}$. As shown in Fig.~\ref{fig:conteggi}, once fitted the total differential
logN-logS, our model predict the correct number of BL-Lacs and
FSRQs without any change in the radio LF. This fact support
the correctness of the initial hypothesis of radio galaxies as parental population
of blazars.
The number ratio of blazars to radio galaxies $\kappa$ can be thought as a measure of the 
beaming factor of the relativistic jet, which in turn is related to
the bulk Lorentz factor $\Gamma$. From $\kappa \sim 1/2\Gamma^2$ we
derive $\Gamma \sim 35$.\\
Fig.~\ref{fig:paola} shows the corrisponding contribution of FSRQs and
BL-Lac to the total \Fermi-LAT EGB. The background intensity is found
to be $I_{\rm{FSRQ}}=4.22 \times 10^{-6} \rm{ph}~
\rm{s^{-1}} \rm{cm^{-2}} \rm{sr^{-1}}$ and  $I_{\rm{BL-Lac}}=2.43 \times 10^{-6} \rm{ph}~
\rm{s^{-1}} \rm{cm^{-2}} \rm{sr^{-1}}$ for FSRQs and BL-Lacs
respectively. The total EGB intensity is therefore  $I=6.65 \times 10^{-6} \rm{ph}~
\rm{s^{-1}} \rm{cm^{-2}} \rm{sr^{-1}}$, corrisponding to $~45\%$ of
the one measured by \Fermi-LAT.\\
From the slope of the FSRQ and BL-Lac component in Fig.~\ref{fig:paola} we can see that the main photon index of
FSRQs and BL-Lacs resulting from our model are in agreement with the
2LAC \cite{Fermi_coll}.\\
We note that blazars fall short to explain the measured EGB at E$<$10 GeV and at E$>$50 GeV. At low energies, the discrepancy can be fully accounted by star-forming galaxies modeled following the recipes by\cite{Stecker}, so that only the last point of the \Fermi-LAT EGB measurement is not reproduced by our model.

\section{Discussion}

We have computed the overall contribution of blazars to the \Fermi-LAT
EGB.

Our model relies on two assumptions: the radio
LF and the blazar SED. In the following we show the difference with
the most reliable works on the contribution of blazars to the
\Fermi-LAT EGB.
\begin{itemize}
\item We use the radio LF \cite{Willott}. Fitting the differential
  logN-logS we obtain just the overall normalization $k$, without any
  change on the bright and faint end of LF.\\
  In previous works \cite{Inoue}\cite{Narumoto} it
  is assumed a LF in X band with three free parameters: the total
  normalization, the amount of bolometric radiation emitted in X-ray
  and the faint end of the X-ray LF. Differently in \cite{Stecker}
  the radio LF computed by (\cite{Dunlop}) is used, 
  changing the faint end to obtain the contribution of FSRQs.
 \item The second main point of our model is represented by the blazar
  SED. We assume that the spectra of blazar is fully taken into
  account by the \emph{blazar sequence} \cite{Fossati}. On the contrary
  in \cite{Stecker, vent} 
it is assumed the blazar spectra as a simple and
  broken power law, respectively. In these works they assume a
  spectral index distribution peaked on the mean spectral index
  resulting from observations. The underlying assumption is that the
  unresolved blazar component has the same index distribution of the
  resolved component. Using the blazar sequence such assumption is not
  necessary because the blazar SEDs are fully determined.
\end{itemize}

\begin{figure}[t]
\centering
\includegraphics[scale=0.43]{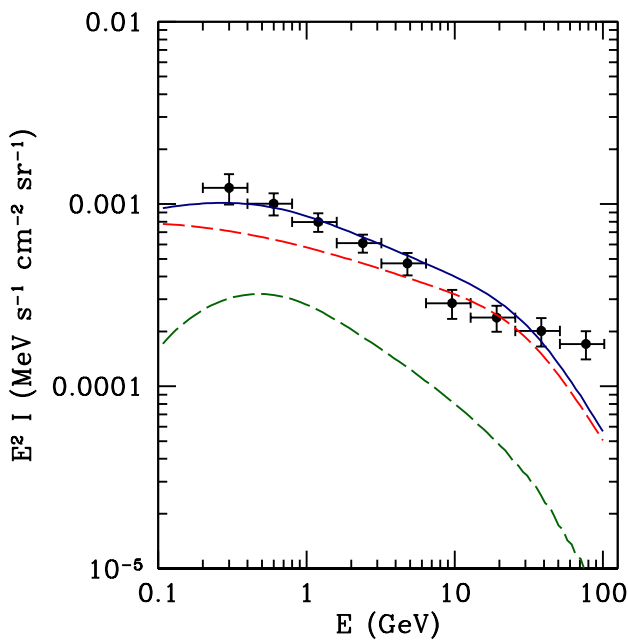}
\caption{The star-forming galaxy component (green line), the total
  blazar component (red line) and the sum (blue line). In black are
  the \Fermi-LAT EGB points.} \label{fig:totale}
\end{figure}

The best fit value of the relative number of blazars with respect to radio galaxies can be translated into a bulk Lorentz factor of the 
relativistic jet $\Gamma \sim 35$, larger than the average value
$\Gamma\sim 15$ estimated in \cite{Ghisellini}. The two values
could be reconciled if blazars commonly show secular $\gamma$-ray
large variability which modulates the 1-year average flux, as recently
proposed in \cite{Ghirlanda}.\\
To be consistent with the \Fermi-LAT points at lower energy, we add
the star-forming component to our blazar model. 
Fitting the \Fermi-LAT EBG with this two component model (Fig.~\ref{fig:totale}), we
constrains the so-called `` star formation efficiency of molecular
hydrogen" $\xi=7.0\times 10^{-10}$yr$^{-1}$,  which we found well within existing, much looser
observational constraints \cite{Leroy}.\\
The point at 80 GeV remains unaccounted by our models calling for an
additional (local or low-z) component. An intriguing possibility is
represented by annihilation of a dark matter particle. We found, as an example, that   
a particle of mass $\simeq 0.5$ TeV and cross section $\langle \sigma v \rangle \simeq 5 \times  10^{-26}$ cm$^3$ s$^{-1}$ 
can easly accomodate the last data point.

\bigskip % extra skip inserted

\begin{thebibliography}{9} 

\bibitem{Fic78} Fichtel, C.~E., 
Simpson, G.~S., \& Thompson, D.~J.\ 1978, baas, 10, 395 

\bibitem{Strong} Strong, A.~W., 
Moskalenko, I.~V., \& Reimer, O.\ 2004, ApJ, 613, 956 

\bibitem{Abdofondo} 
Abdo A.~A., et al., 2010a, PhRvL, 104, 101101 



\bibitem{Fermi_coll} The Fermi-LAT collaboration, 
2011, arXiv, arXiv:1108.1420 

\bibitem{Dermer} Dermer 
C.~D., 2007, ApJ, 659, 958 
\bibitem{Inoue} Inoue, Y., \& Totani, T.\ 2009, ApJ, 702, 523 


\bibitem{Stecker} Stecker F.~W., Venters T.~M., 2010, arXiv, arXiv:1012.3678 

\bibitem{vent} Venters, T.~M., \& Pavlidou, V.\ 2011, ApJ  737, 80 



\bibitem{Willott} Willott C.~J., Rawlings S., Blundell 
K.~M., Lacy M., Eales S.~A., 2001, MNRAS, 322, 536 

\bibitem{Donato} Donato D., Ghisellini G., Tagliaferri G., Fossati G., 2001, A\&A, 375, 739 


\bibitem{Fossati} Fossati G., Maraschi L., Celotti A., 
Comastri A., Ghisellini G., 1998, MNRAS, 299, 433
 

\bibitem{AbdoLF} 
Abdo A.~A., et al., 2009, ApJ, 700, 597 







\bibitem{Finke} Finke J.~D., Razzaque S., Dermer C.~D., 2010, ApJ, 712, 238 

\bibitem{Narumoto} Narumoto, T., \& Totani, T.\ 2006, ApJ, 643, 81 


\bibitem{Abdocount} 
Abdo A.~A., et al., 2010, ApJ, 720, 43

\bibitem{Dunlop} Dunlop J.~S., Peacock J.~A., 1990, MNRAS, 247, 19 


\bibitem{Ghirlanda} Ghirlanda G., 
Ghisellini G., Tavecchio F., Foschini L., Bonnoli G., 2011, MNRAS, 413, 852 



\bibitem{Ghisellini} Ghisellini G., et 
al., 2010, MNRAS, 405, 387 


\bibitem{Leroy} 
Leroy A.~K., Walter F., Brinks E., Bigiel F., de Blok W.~J.~G., Madore B., 
Thornley M.~D., 2008, AJ, 136, 2782 




\end{thebibliography}

\end{document}